\documentclass[pra,amssymb]{revtex4}

\usepackage{appendix}
\usepackage{graphicx}
\usepackage{amssymb,amsmath} 
\usepackage{epsfig}
\usepackage[utf8]{inputenc}
\usepackage[T1]{fontenc}

\def\B#1{\!\left(#1\right)}
\def\BB#1{\!\left[#1\right]}

\def\p2Var{\partial^2_\eta{\rm Var}}

\def\bg{\begin{equation}\begin{gathered}}
\def\eg{\end{gathered}\end{equation}}

\def\be{\begin{equation}}
\def\ee{\end{equation}}

\def\bx{\begin{equation}\begin{aligned}}
\def\ex{\end{aligned}\end{equation}}

\def\bee{\begin{equation*}}
\def\eee{\end{equation*}}

\def\pref{0.65}
\def\prefmax{0.65}

\begin{document}

\title{Spatial Kibble-Zurek mechanism through susceptibilities: the  inhomogeneous quantum Ising model case}

\author{Mateusz \L\k{a}cki$^{1,2}$ and  Bogdan Damski$^{3}$} 
\affiliation{
\mbox{$^{1}$ Institute for Theoretical Physics, University of Innsbruck, A-6020 Innsbruck, Austria}		
\mbox{$^{2}$ Institute for Quantum Optics and Quantum Information of the Austrian Academy of Sciences, A-6020 Innsbruck, Austria}
\mbox{$^{3}$ Jagiellonian University, Institute of Physics, {\L}ojasiewicza 11, 30-348 Krak\'ow, Poland}
}
\begin{abstract}
We study the quantum Ising model in the transverse inhomogeneous magnetic field. 
Such a system can be approached  numerically through exact diagonalization and analytically through the
renormalization group techniques. Basic insights into its physics, however, 
can be obtained by adopting the Kibble-Zurek theory of
non-equilibrium phase transitions to description of spatially inhomogeneous  systems at equilibrium. 
We employ all these approaches and focus on derivatives of 
longitudinal and transverse magnetizations, which have extrema near the
critical point. We discuss how these extrema can be used for locating the
critical point and for verification of the Kibble-Zurek scaling predictions in
the spatial quench. 
\end{abstract}
\maketitle

\section{Introduction}
\label{introduction_sec}
The Kibble-Zurek (KZ) mechanism provides basic insights into dynamics of 
non-equilibrium phase transitions. It was introduced by Kibble to describe  
 phase transitions  of the early Universe \cite{Kibble1976} 
and then it was extended to condensed matter systems 
by Zurek \cite{Zurek1985}. Over the years, the KZ mechanism  has been studied in numerous  systems undergoing classical 
phase transitions \cite{KZclassRev}.
Three (two) decades after the influential  work of Kibble (Zurek), the KZ mechanism was applied to quantum systems 
\cite{BDPRL2005,DornerPRL2005,JacekPRL2005,PolkovnikovPRB2005}, and then it was 
vigorously studied in ubiquitous models undergoing non-equilibrium quantum phase transitions
\cite{JacekAdv2010,PolkovnikovRMP2011,Dutta2015}. 
We will focus our attention below on quantum systems
only.

The  idea behind the KZ studies of non-equilibrium quantum dynamics is simple. 
Prepare the system in a ground  state far away from the critical point, 
drive it from one phase to another by the global change of the external field, 
and use its equilibrium properties to predict its  excitation due to crossing of the critical point. 

There are three generic stages of such time evolutions. Initially, the excitation gap is large enough to ensure
nearly adiabatic dynamics. Therefore, the system is in the instantaneous ground
state as it is driven towards the critical point. It then goes out of equilibrium near  the critical point
because the excitation gap vanishes at the critical point, and so hardly any energy
is required to excite the system there. The
excitations are produced until the gap in the excitation spectrum becomes large enough
on the other side of the transition. Next, the second adiabatic stage takes
place, where the probability of finding the system in the  instantaneous
ground state is  again nearly constant, but this time different from one due to
the system excitation.
It is then of interest to estimate the size of the
crossover region between the two adiabatic limits and to use it to link the 
dynamics of the  system  to the rate of driving.  

More quantitatively, let's assume that the external field $g(t)$ driving the transition is changed linearly
in time
\bee
g(t)=g_c+\frac{t-t_c}{\tau_Q}, 
\eee
where $g_c$ is the location of the critical point and $\tau_Q$ is the quench time controlling the driving rate. 
The critical point is crossed at time $t_c$. The location of the crossover from the adiabatic to
non-adiabatic dynamics can be estimated  from  equation \cite{BDPRL2005,DornerPRL2005,JacekPRL2005}
\be
\frac{\hbar}{\Delta\BB{g(t_c\pm\hat t\,)}}\sim \hat t,
\label{delta_that}
\ee
which expresses the  observation that evolution ceases to be adiabatic when
the reaction time of the system, which is proportional to the inverse of the excitation
gap $\Delta$, becomes comparable to the time $\hat t$ that is left to reaching the
critical point. In other words, the crossover from the adiabatic to
non-adiabatic dynamics  takes place at the distance $\hat g=|g_c-g(t_c\pm\hat t\,)|=\hat t/\tau_Q$ from the critical point. 
Taking the standard scaling relation  
\bee
\Delta\sim|g-g_c|^{z\nu},
\eee
where $z$ and $\nu$ are the dynamical and correlation-length exponents,
respectively, one easily finds characteristic non-equilibrium time, field, and
length scales
\bee
\hat t\sim \tau_Q^{z\nu/(1+z\nu)},  \ \ \hat g\sim\tau_Q^{-1/(1+z\nu)}, \ \ 
\hat\xi\sim\tau_Q^{\nu/(1+z\nu)}, 
\eee
where  $\hat\xi$ is the correlation length at the crossover between adiabatic and non-adiabatic regimes. 
It is assumed here that the correlation length behaves near the critical point as
\be
\xi(g)\sim\frac{1}{|g-g_c|^\nu}.
\label{xii}
\ee
The non-equilibrium state of the system after the quench depends on $\hat t$, $\hat g$, and $\hat\xi$,
which is the essence of the Kibble-Zurek theory.

The spatial analog of the above time quench is obtained by replacing  the driving in the 
time domain by the driving in the spatial domain \cite{TurbanJPA2007,DornerPRS2008}. 
This is achieved by making the 
field position-dependent only 
\be
g(x)=g_c+\frac{x-x_c}{\lambda_Q},
\label{gx}
\ee
where $\lambda_Q$ is the rate of the
spatial quench and $x$ is the distance within the system (we assume that the system is one dimensional). 
The field reaches then the critical value $g_c$ at $x_c$,
which we will call below a spatial critical point.

It is now convenient to introduce the local homogeneous approximation. 
Let   $O$ be an observable of interest, which  in a ground state of the
homogeneous system, exposed to the  external field $g$, equals $O_H(g)$. 
The local  homogeneous approximation assumes that in the ground state of the inhomogeneous system 
$O[g(x)]$ is accurately approximated by $O_H[g(x)]$.

We expect that far away from the spatial critical point the local homogeneous approximation holds.
This is analogous to the above-invoked statement that the system undergoes
adiabatic dynamics far away from the critical point during the time quench.
Around the spatial critical point, however, the local homogeneous approximation
has to break down just as the adiabatic approximation breaks 
near the critical point during the time quench. The crossover between the two regimes is
estimated from the spatial analog of (\ref{delta_that})
\bee
\xi[g(x_c\pm\hat x)]\sim\hat x,
\eee
whose solution provides the driving-induced length and field scales
\be
\hat x  \sim \lambda_Q^{\nu/(1+\nu)}, \ \ \hat g\sim \lambda_Q^{-1/(1+\nu)},
\label{xhatghat}
\ee
where now $\hat g=|g_c-g(x_c\pm\hat x)|=\hat x/\lambda_Q$  \cite{DornerPRS2008}.  The spatial quench has been studied in 
the quantum Ising model \cite{DornerPRS2008,TurbanJPA2007,TurbanJStat2009},  the spin-1 Bose-Einstein condensate 
\cite{BDNJP2009}, the XY model \cite{CampostriniPRA2010}, and the  XXZ model \cite{SuzukiPRB2015}.
We will look at it in more detail in the following sections.

Finally, we mention that besides the time and space quenches,  the space-time quenches have been also studied in the
context of the KZ mechanism. The parameter inducing the transition in such quenches is both space- and
time-dependent \cite{JacekMarekNJP2010,ColluraPRA2011,MarekAdolfoNJP2016}.

\section{Spatial Kibble-Zurek mechanism}
\label{spatial_sec}
We have discussed so far how the size of the crossover region around the critical point 
scales with the quench rate $\lambda_Q$ (\ref{xhatghat}). 
Suppose now that we know $O[g(x)]$ from either theory or experiment.
To verify such a scaling prediction {\it directly}, one needs to
study deviations from the local homogeneous approximation
\be
O[g(x)]-O_H[g(x)]
\label{diff}
\ee
as a function of the distance $|x-x_c|$ from the spatial critical point.
This means that one needs to know {\it a priori} the properties of a homogeneous
system, $O_H(g)$ and $g_c$, to determine the  basic scaling predictions in the spatial quench (\ref{xhatghat}). 
While such data   can be readily available in exactly solvable systems, it  can be difficult to obtain in other  
systems undergoing a quantum phase transition
(e.g. due to hard-to-eliminate  inhomogeneities such as those
induced by ubiquitous external trapping in cold atom experiments \cite{Lewenstein,BlochRMP2008}).

A different  strategy to verify the scaling predictions (\ref{xhatghat}), and perhaps more importantly 
to get insights into the behavior of our observable near the spatial critical point,
is suggested by the scaling ansatz approach. To discuss it, we assume that 
the observable is known to have algebraic singularity near the
critical point of a homogeneous system
\be
O_H(g)\sim \frac{1}{|g-g_c|^\gamma},
\label{OHsing}
\ee
and write the scaling ansatz as 
\be
O[g(x)]=\hat x^{\gamma/\nu}f\B{\frac{x-x_c}{\hat x}},
\label{ansatz}
\ee
where $f$ is a scaling function that does not have to be known {\it a priori}; one only assumes that $f(0)\neq0$.  
Such an ansatz was first derived through the renormalization group (RG) study without invoking the KZ
arguments in an elegant reference  \cite{TurbanJPA2007}. 
Several remarks are in order now.

First, perhaps the best insight into this plausible ansatz is provided by the
finite-size scaling theory,
and it is worth to
 present it here
\cite{TurbanJPA2007,Cardy1}. Suppose the observable scales near the critical point of the infinite  homogeneous 
 system as (\ref{OHsing}). Such an observable in a finite system
 may depend on three  length scales: the correlation length $\xi$, the system size $N$, and some system-specific 
microscopic length scale. The finite-size scaling hypothesis assumes that near the critical 
point the  microscopic length scale does not matter and so the observable should depend  
on either $\xi$ or $N$ and the ratio between these two  length
scales. Using (\ref{xii}) and noting that the right-hand-side (RHS) of (\ref{OHsing}) is proportional 
to $\xi^{\gamma/\nu}$, the following ansatz can be 
proposed 
\be
\left.O_H(g)\right|_{N<\infty}=\xi^{\gamma/\nu}
\phi\B{\frac{N}{\xi}}=N^{\gamma/\nu}\tilde\phi\B{\frac{N}{\xi}},
\label{fs_ansatz}
\ee
where it is assumed that $\phi(y\to0)\sim y^{\gamma/\nu}$ to cancel the singularity at the critical point
when $g\to g_c$, and $\tilde\phi(y\to\infty)\sim y^{-\gamma/\nu}$ to recover singularity (\ref{OHsing})
when $N\to\infty$. Other limits of these functions can be obtained from the
mapping $\phi(y)=y^{\gamma/\nu}\tilde\phi(y)$. Ansatz (\ref{ansatz}) is
now obtained by replacing the system size $N$ in (\ref{fs_ansatz}) by $\hat x$ from (\ref{xhatghat})
and then by  identifying $\tilde\phi(y)$ with $f(\pm y^{1/\nu})$ \cite{TurbanJPA2007}. 
Therefore,  in the light of the finite-size scaling theory,  ansatz (\ref{ansatz}) supports the view
that near the spatial critical point the inhomogeneous system behaves as if it would have a finite
length given by the size $\hat x$ of the crossover region.

Second, such an  ansatz regularizes  observable $O$ near the
critical point by removing its  sharp singularities associated with the quantum
critical point (divergence when $\gamma>0$ or zero when $\gamma<0$). 
Indeed, from (\ref{ansatz}) we  immediately obtain the scaling  of our
observable near the spatial critical point
\be
O(g_c)\sim \lambda_Q^{\gamma/(1+\nu)}.
\label{Ogc}
\ee 
For $\gamma>0$ ($\gamma<0$) this relation shows how fast divergence (zero) is approached when $\lambda_Q\to\infty$.

Third, the  usefulness of  ansatz (\ref{ansatz}) can be also appreciated by looking at 
\be
\lambda_Q^{-\gamma/(1+\nu)} O[g(x)]  =  f\B{\frac{x-x_c}{\hat x}}.
\label{resc}
\ee
If we now know  $\gamma$, $\nu$ and  $g_c$,
we can plot the left-hand-side  of (\ref{resc}) as a function of $(x-x_c)/{\hat x}=(g-g_c)/\hat g$ and
find out the region around the critical point, where the data 
obtained for different $\lambda_Q$'s collapse. This  would indirectly verify
scaling predictions (\ref{xhatghat}). Such an approach
eliminates the need for knowing the  proportionality constant in (\ref{OHsing}) that would be
crucial in the studies based on difference (\ref{diff}).
It still requires knowledge of the critical exponents $\gamma$ and $\nu$ and the location of the
critical point $g_c$. It is thus of great interest to be able to extract out of 
available data for $O[g(x)]$ at least some of these quantities before
studying  the collapse suggested by (\ref{resc}).

Fourth, ansatz (\ref{ansatz}) is analogous to the scaling ansatz frequently discussed in the context of 
time quenches, where it was found that the non-equilibrium dynamics of physical observables depends 
on the rescaled time $t/\hat t$ and distance $x/\hat\xi$. The latest
take on this plausible observation can be found in \cite{KolodrubetzPRL2012,SondhiPRB2012,FrancuzPRB2016} focusing on 
quantum phase transitions. For the preceding work on the dynamics of classical phase transitions pointing 
out in the same direction see e.g.  \cite{BDPRL2010}.

If we now look at (\ref{OHsing}--\ref{resc}), 
we immediately see that it would be useful to know the location of the critical point.
The position of the critical point can be extracted out of the data 
for the inhomogeneous system by studying  the quantity $O$ 
that is  divergent at the critical point of the infinite homogeneous system, $O_H(g_c)=\pm\infty$.
We will call below such a quantity susceptibility.   
Its singularity will be turned in an inhomogeneous field $g(x)$ into an extremum,
which can be used for  finding the critical
point.

The goal of this paper is to study susceptibilities in the inhomogeneous Ising chain 
to see how accurately one may use them to extract the position of the critical point 
and to verify KZ scaling relations in a spatial quench. Sec. \ref{ising_sec}
and  Appendix provide basic information about the  Ising model.
Sec. \ref{longitudinal_sec} is devoted to studies of the derivative of
longitudinal magnetization, which 
has algebraic singularity in a homogeneous system. As longitudinal
magnetization in the inhomogeneous Ising model was studied in \cite{TurbanJPA2007}, 
these results provide a complementary
look at its behavior. Sec. \ref{transverse_sec}
focuses on transverse magnetization, which in the homogeneous system has logarithmic singularity
making ansatz (\ref{ansatz}) inapplicable.  We argue  that the proper
scaling ansatz capturing behavior of the derivative of  transverse magnetization
is highly non-trivial. The results of this paper are briefly summarized in
Sec. \ref{summary_sec}.

\section{Ising model}
\label{ising_sec}
We study the one-dimensional quantum  Ising model in the transverse
inhomogeneous field, whose Hamiltonian reads
\be
\hat H = -\sum_{i=1}^{N-1} \sigma^x_i\sigma^x_{i+1} - \sum_{i=1}^N g_i \sigma^z_i,
\label{HIsing}
\ee
where $N$ is the number of spins and  open boundary conditions  are assumed.

When the system is homogeneous, $g_i=g=\text{const}$, we have the quantum
critical point at $g_c=1$. The properties of the homogeneous Ising model have been
extensively studied over the years, see e.g. the following popular references 
\cite{Lieb1961,Pfeuty}.  For $0\le g<1$ the system is in the
ferromagnetic phase, while for $g>1$ it is in the paramagnetic phase (we
assume $g\ge0$ everywhere in this work). Transverse magnetization is  given   by \cite{Pfeuty}
\be
S^z_H = \frac{g-1}{g\pi}K\left(\frac{2\sqrt{g}}{1+g}\right)+\frac{g+1}{g\pi}E\left(\frac{2\sqrt{g}}{1+g}\right).
\label{Sz_homo}
\ee
The complete elliptic integrals $K$ and $E$ as well as the origin of this standard formula 
are discussed in   Appendix. While such $S^z_H$ is continuous and finite at the critical point,
its derivative over $g$  is logarithmically divergent at the critical point. Using  expressions
from Appendix, one  finds that  very close to the critical point $dS_H^z/dg$ can be approximated by 
\be
\text{const} -\frac{1}{\pi}\ln|g-1|.
\label{1_2pi}
\ee

Longitudinal magnetization in the homogeneous system is given by \cite{Pfeuty}
\be
S^x_H= \theta(1-g)\B{1-g^2}^{1/8},
\label{Sx_homo}
\ee
where $\theta$ is the Heaviside step function. This implies that for $g\neq1$
\be
\frac{dS^x_H}{dg}=-\frac{g\theta(1-g)}{4(1-g^2)^{7/8}},
\label{dSx_homo}
\ee
which is singular at the critical point. To close the discussion of the homogeneous Ising model in the
transverse field,  we mention that its correlation-length critical exponent $\nu$  equals one \cite{Barouch1971}.

\begin{figure}[t]
\includegraphics[width=\pref\columnwidth,clip=true]{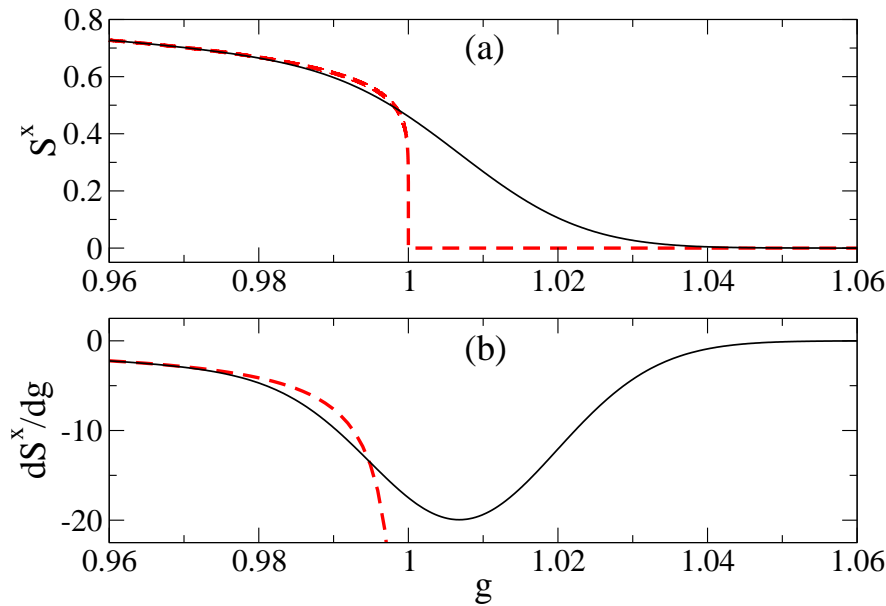}
\caption{Longitudinal magnetization and its derivative. 
In both panels results for inhomogeneous (homogeneous) systems are denoted by  the black solid (red dashed) line.
The homogeneous results, $S^x_H$ and $dS^x_H/dg$,  are given by
(\ref{Sx_homo}) and (\ref{dSx_homo}), respectively.
The inhomogeneous field is given by (\ref{gi}) with $\lambda_Q=5000$.  }
\label{fig1}
\end{figure}

We  assume from now on that the system is placed in the inhomogeneous
field that  linearly varies with the spin site index $i$
\be
g_i = g_c + \frac{i-N/2}{\lambda_Q}.
\label{gi}
\ee
We follow reference \cite{YoungPRB1997} to diagonalize  Hamiltonian (\ref{HIsing}).
On the technical level, this implies that 
we have to find a full spectrum of the $2N\times2N$ real symmetric matrices. 
This requirement limits system sizes that can be 
efficiently simulated to about $10^4$ spins. We have chosen in all our calculations 
system sizes so large, that the finite-size corrections 
are negligible in all the plots that we discuss. 
Therefore, we do not report
system sizes in Figs. \ref{fig1}--\ref{fig7}. 
The system sizes needed to achieve such a
limit can be estimated by requiring that 
\be
N\gg\hat x,
\label{Nlarge}
\ee
which in our calculations, where  $\lambda_Q<10^6$, implies that $N\gg10^3$.
We have used $N$ up to $8192$ for the largest $\lambda_Q$'s. 

The eigenvectors of the above-mentioned  matrices
are then used to construct the Bogolubov modes diagonalizing the Hamiltonian. They
are then  employed  to compute 
$S^z$ and $S^x$ magnetizations. The calculation of the former is
straightforward, while the computation of the latter is a bit tricky for the
following reason. Due to the lack of the symmetry-breaking longitudinal field 
in  Hamiltonian (\ref{HIsing}), the expectation value of the operator 
$\sigma^x_i$ is exactly zero in any of its eigenstates. This difficulty can be solved,
without adding the symmetry-breaking field, by computation of the matrix
element of the $\sigma^x_i$ operator between the two lowest-energy
eigenstates of Hamiltonian (\ref{HIsing}); see \cite{TurbanJPA2007} for
details. Such a matrix element can be evaluated with the help of the Wick
theorem \cite{Wick}.

Having computed magnetizations $S^{x,z}(g_i)$, where $i$ enumerates lattice sites, 
we calculate their numerical derivatives. Some care has to be exercised here.
Adopting the symmetric  expression for the derivative, 
we get 
\bee
\frac{dS^{x,z}}{dg}(g_{i+1/2})\approx\frac{S^{x,z}(g_{i+1})-S^{x,z}(g_i)}{g_{i+1}-g_i}.
\eee
We locate then the lattice sites, where the   
derivatives attain extremal values. Next, we fit a polynomial of $6$th order to the data
around such lattice sites, and associate the extrema of the fitted polynomials
with the extrema of $dS^{x,z}/dg$.

\section{Longitudinal magnetization}
\label{longitudinal_sec}

Longitudinal magnetization and its derivative are  plotted in Fig. \ref{fig1}. 
It is clear from this plot that magnetization $S^x$, unlike its homogeneous
counterpart $S^x_H$, is featureless around the critical point. It smoothly
decreases reaching   zero far away from the critical point on the paramagnetic side of the
phase diagram (Fig. \ref{fig1}a). 
The most efficient way to estimate the position of the critical point from
such data is to compute the derivative of $S^x$, which exhibits the minimum  near the critical point 
in the paramagnetic phase (Fig. \ref{fig1}b). 
It is therefore quite interesting to find out how accurately one may
extrapolate the position of the critical point from such data. 

Since the singularity of $dS_H^x/dg$ is algebraic (\ref{dSx_homo}), we use 
scaling ansatz (\ref{ansatz}) to study the extremum of  $dS^x/dg$. It then immediately follows that this extremum
should be located at $g_x^*$ such that 
\be
|g_x^*-g_c|\sim \hat g\sim \lambda_Q^{-1/(1+\nu)}.\\
\label{gstar_scal}
\ee
Moreover, we also expect from (\ref{ansatz}) that 
\be
\frac{dS^x}{dg}(g_x^*)\sim{\hat x}^{\gamma/\nu}\sim \lambda_Q^{\gamma/(1+\nu)}.
\label{ostar_scal}
\ee
These scaling predictions will be verified below. Before moving on, however, 
we would like to note that the scaling relation (\ref{ostar_scal})
involves the exponent $\gamma=7/8$, which  is defined only on the ferromagnetic side of the
transition (\ref{dSx_homo}). Therefore, it should not be taken for granted
that the scaling of $dS^x/dg$ at the minimum, which is located in the paramagnetic phase,
is given by relation (\ref{ostar_scal}).

\begin{figure}[t]
\includegraphics[width=\pref\columnwidth,clip=true]{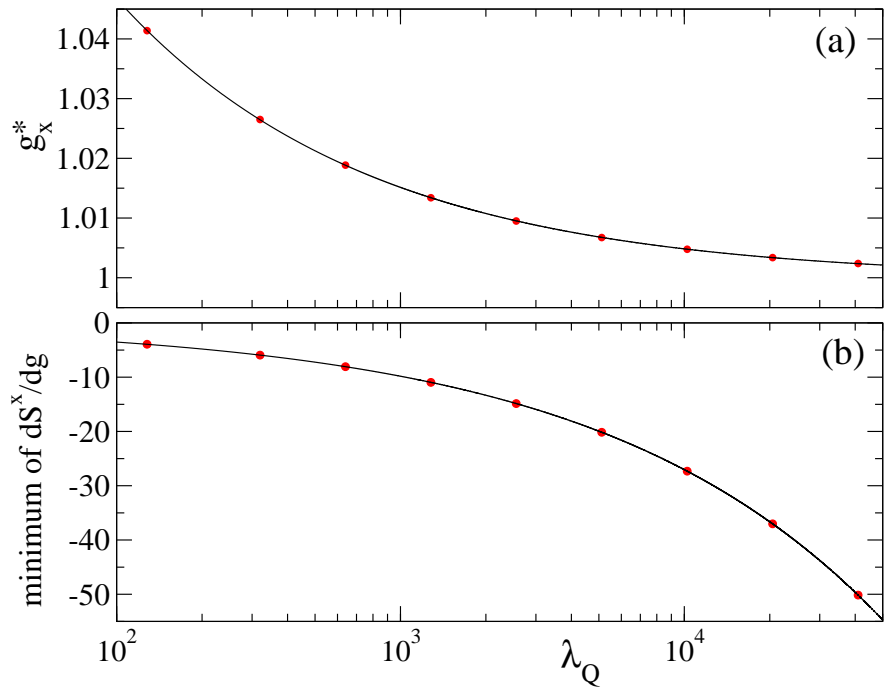}
\caption{Scaling properties of the minimum of $dS^x/dg$. Panel (a): 
position $g^*_x$ of the minimum. 
Red dots provide numerical data. The black line is the non-linear fit yielding
$g^*_x=0.99985(3)+0.440(2)\lambda_Q^{-0.487(1)}$.
Panel (b): minimum of $dS^x/dg$. 
Red dots provide numerical data, while the  black line is the non-linear fit: minimum of 
$dS^x/dg=0.095(3) -0.4845(4)\lambda_Q^{0.43705(7)}$. 
The fitted coefficients and their standard errors, both here and in other plots in this work, come from  
NonlinearModelFit function from \cite{Mathematica}. We report everywhere in
this paper  one standard error in the brackets next to the fitted 
coefficients.
}
\label{fig2}
\end{figure}

The scaling of  $g^*_x$  is presented in Fig.
\ref{fig2}a. We fit there  $g^*_x$ with 
\be
 a + b\lambda_Q^{c}.
\label{fit}
\ee
The parameter $a$ extrapolates the position of the critical point to
the infinite $\lambda_Q$ limit, where the system becomes effectively
homogeneous ($c<0$). We obtain $a\approx0.99985$, which proves that 
susceptibility $dS^x/dg$ in the spatially inhomogeneous system can be
efficiently used for finding the critical point. 

The parameter $c$, according to (\ref{gstar_scal}), should be
compared to $-1/(1+\nu)=-1/2$. 
The fitted exponent  $c\approx-0.487$  is in a good agreement 
with the spatial KZ scaling arguments employed in ansatz (\ref{ansatz}).
Its actual value depends on the range of $\lambda_Q$'s used for the fit. We
used $\lambda_Q\in(10^2,5\cdot10^5)$ to obtain such a value. 
Increase of the lowest $\lambda_Q$'s used for the fit, brings the exponent $c$
closer to the expected value of $-1/2$. This is in agreement with the
expectation that KZ scaling predictions work best in the limit of $\lambda_Q\to\infty$. 
The same happens in time quenches, where the KZ predictions best fit the non-equilibrium 
data in the limit of $\tau_Q\to\infty$ as long as finite-size effects are
negligible, which happens for \cite{JacekPRL2005}
\be
N\gg\hat\xi.
\label{Ntlarge}
\ee
Note the similarity between the conditions for irrelevance of  finite-size effects in 
spatial (\ref{Nlarge}) 
and time  (\ref{Ntlarge})
quenches.

The scaling of the value of $dS^x/dg$ at the minimum is illustrated in Fig.
\ref{fig2}b. Fit (\ref{fit}) to the numerical data provides the exponent 
$c\approx0.43705$, which is in excellent agreement with the expected 
$\gamma/(1+\nu)=7/16=0.4375$ from (\ref{ostar_scal}). This happens  despite the above-mentioned 
problems with assigning the value to the exponent $\gamma$ on the paramagnetic
side of the transition.

Looking at these results, we see an immediate analogy to the ones  that are
obtained in homogeneous but finite-size systems \cite{Cardy1,Cardy2}. In such systems
susceptibilities also exhibit an extremum near the critical point. Such an
extremum  moves towards the critical point when the system size increases.
The extrapolation of its  position to the thermodynamic limit  provides the critical
point. The algebraic in the system size scaling of its distance and value 
encodes the critical exponents $\nu$ and $\gamma$, respectively.

\section{Transverse magnetization}
\label{transverse_sec}

\begin{figure}[t]
\includegraphics[width=\pref\columnwidth,clip=true]{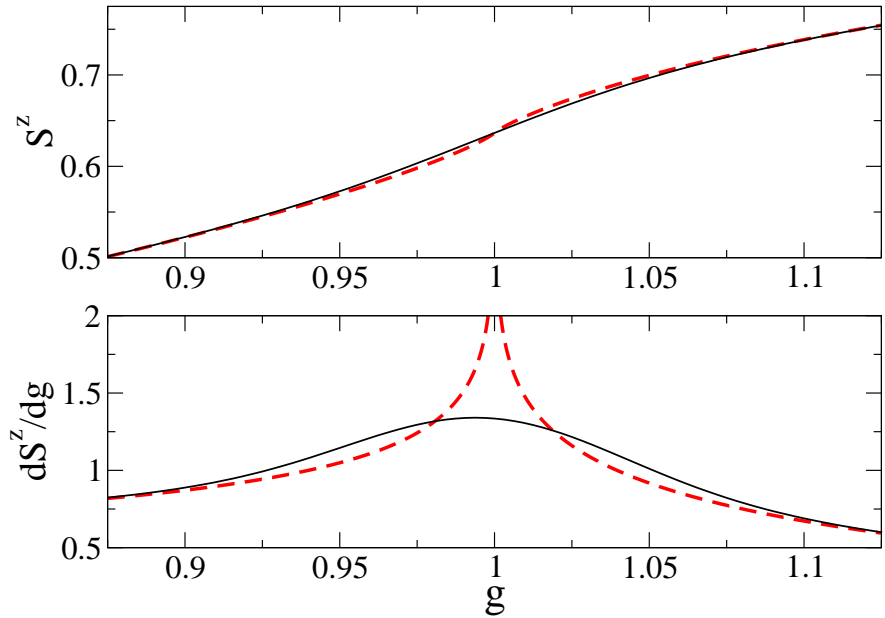}
\caption{Transverse magnetization and its derivative. 
In both panels results for inhomogeneous (homogeneous) systems are denoted by  the solid black (dashed red) line.
The homogeneous results $S^z_H$ and $dS_H^z/dg$ are given by (\ref{Sz_homo}) and
(\ref{dSzz}), respectively.
The inhomogeneous results come from numerics done for  $\lambda_Q=320$. }
\label{fig3}
\end{figure}

Transverse magnetization $S^z$ and its derivative $dS^z/dg$ are
plotted in Fig. \ref{fig3}.  
It is interesting to note that  even in the thermodynamically-large 
homogeneous system  $S^z_H$, unlike $S^x_H$,  does not reveal where the critical point is
until we compute its derivative. 
The derivative of transverse magnetization 
is quite different from the derivative of longitudinal magnetization in
both homogeneous and spatially driven systems.  Instead of
a minimum  on the paramagnetic side of the transition, there is a maximum of $dS^z/dg$
in the ferromagnetic  phase. $dS^z_H/dg$ is non-zero on both sides of the critical point
unlike $dS^x_H/dg$. Most importantly, however, $dS^z_H/dg$  is logarithmically
rather than algebraically divergent at the critical point.

Therefore, we need to replace ansatz (\ref{ansatz}) to properly take care of the logarithmic singularity (\ref{1_2pi})
at the critical point of the homogeneous system.
To  derive the appropriate ansatz, we adopt to our observable the RG approach developed in \cite{TurbanJPA2007}.
We start by noting that relation (\ref{xii}) implies that  under the change of
the length scale by a factor $b$, $g$ is mapped to $g'$ in a homogeneous system such that \cite{Cardy2}
\be
g'-g_c= (g-g_c)b^{1/\nu}.
\label{gp_gc}
\ee
Thus, if $O_H(g-g_c)$ is given by (\ref{1_2pi}), then the scale transformation leads to
\be
O_H(g'-g_c)= O_H(g-g_c)-\frac{\ln b}{\pi\nu}.
\label{OHmap}
\ee
The scale transformation in the inhomogeneous system results in  
\bee
x'-x_c= (x-x_c)b^{-1}, \ \ \lambda_Q'= \lambda_Q b^{-(1+\nu)/\nu},
\eee
where the first relation is obvious, while the second one is discussed in
\cite{TurbanJPA2007}. Next, we assume that the transformation law (\ref{OHmap}) applies also to 
observable $O$ in an inhomogeneous system sufficiently close to  the critical point. This allows us
to write 
\be
O(x-x_c,\lambda_Q)= O\B{(x-x_c)b^{-1},\lambda_Q
b^{-(1+\nu)/\nu}}+\frac{\ln{b}}{\pi\nu}.
\label{Ob}
\ee
One way to find the scaling solution of  that equation  is to notice that its left-hand-side is
independent of $b$ and so its derivative with respect to $b$ is zero. This
allows us to write the differential equation for $O$
\bee
A\frac{\partial O(A,B)}{\partial A}+\frac{\nu+1}{\nu}B\frac{\partial
O(A,B)}{\partial B}=\frac{1}{\pi\nu}.
\eee
Its regular at $x=x_c$ solution reads 
\be
O(A,B)=f\B{AB^{-\nu/(1+\nu)}}+\frac{\ln B}{\pi(1+\nu)},
\label{OAB}
\ee
where $f$ is an analytic function at the critical
point because we do not anticipate criticality-induced singularities in an
inhomogeneous system.  Alternatively, one may explore the
independence of the left-hand-side of (\ref{Ob}) on the scale factor $b$
by putting $b\sim\lambda_Q^{\nu/(1+\nu)}$ into the RHS of
that equation. This leads to the same solution as (\ref{OAB}). As this is a
quicker approach, we will follow it in the subsequent calculations.  

\begin{figure}[t]
\includegraphics[width=\prefmax\columnwidth,clip=true]{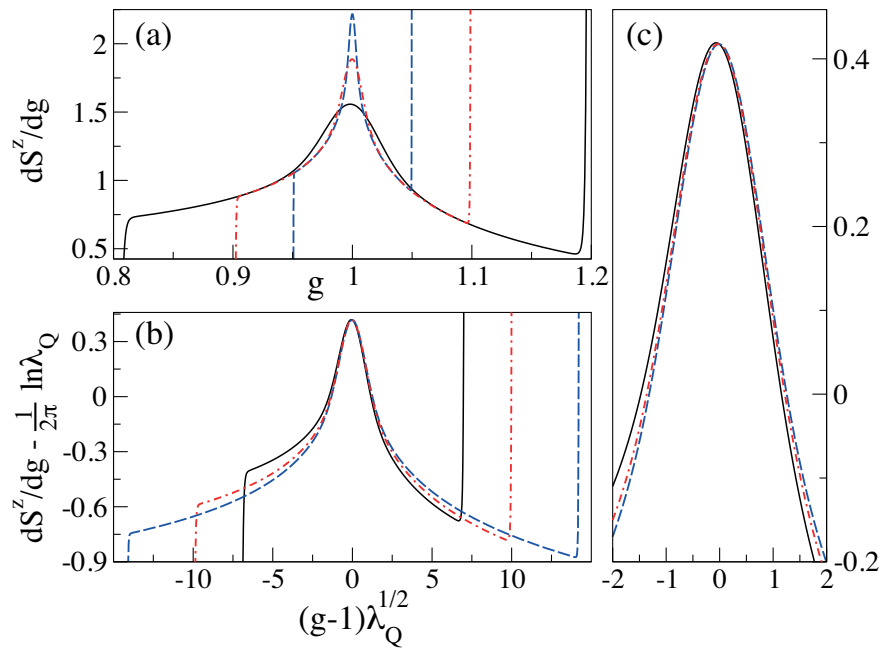}
\caption{Test of ansatz (\ref{ansatzNEW}). 
Panel (a): $dS^z/dg$ before rescaling for $\lambda_Q$ equal
to $1280$ (solid black), $10240$ (dashed-dotted red), and $81910$ (dashed blue).
Panel (b): the rescaled version of the panel (a).   Panel (c): the same as in
the panel (b) but focused around the maximum. The sharp increase of curves in
panels (a) and (b) is caused by open boundary conditions, which 
are imposed at $g$ equal to $1\pm0.2$ ($\lambda_Q=1280$), 
$1\pm0.1$ ($\lambda_Q=10240$), and $1\pm0.05$ ($\lambda_Q=81910$).
}
\label{fig4}
\end{figure}

\begin{figure}[t]
\includegraphics[width=\pref\columnwidth,clip=true]{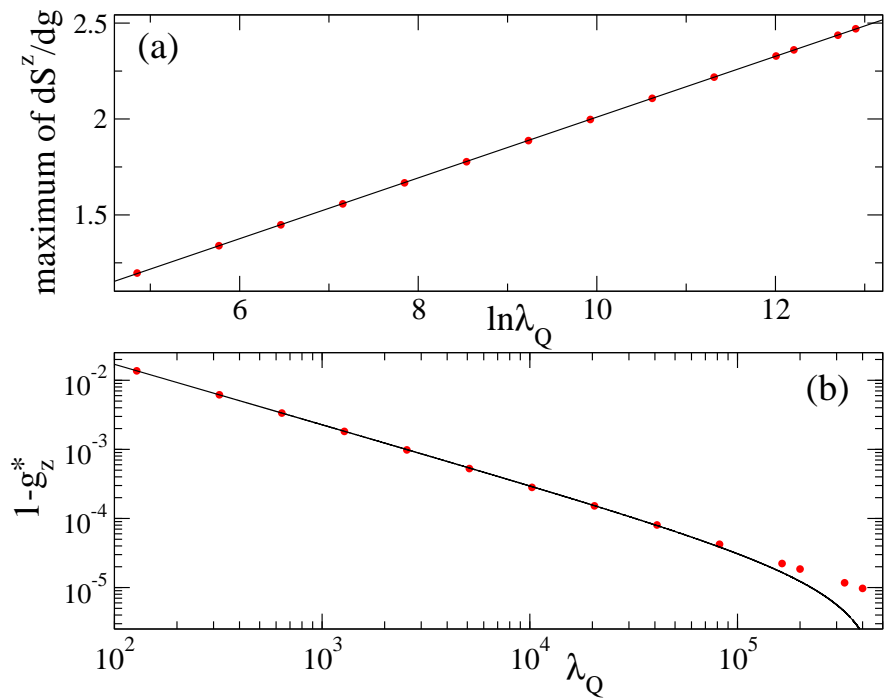}
\caption{Scaling properties of the maximum of $dS^z/dg$.
Panel (a):  Red dots provide numerical data, while  the black line is the 
linear fit:  maximum of $dS^z/dg=0.425(1) + 0.1584(1)\ln\lambda_Q$. 
Panel (b): the position $g^*_z$ of the maximum. Red dots provide numerical data, while the black line comes from  the
non-linear fit $g^*_z=1.00001(3)-0.951(6)\lambda_Q^{-0.873(1)}$. 
}
\label{fig5}
\end{figure}

In the end, we set $\nu=1$, $O=dS^z/dg$, and use (\ref{gx}) and
(\ref{xhatghat}) to arrive at 
\be
\frac{dS^z}{dg}= f\B{\frac{g-g_c}{\hat g}}+\frac{\ln\lambda_Q}{2\pi}.
\label{ansatzNEW}
\ee
Such an ansatz is quite different from the one provided in (\ref{ansatz})
and we test it below. Our findings are 
illustrated in Figs. \ref{fig4} and \ref{fig5}. 

To start discussion of ansatz (\ref{ansatzNEW}), we note that it contains the
KZ-related physics in two places. 
First, the argument of the scaling function $f$  depends in the expected way 
on the  KZ driving-induced field-scale $\hat g\sim1/\sqrt{\lambda_Q}$.
Second, the logarithmic in $\lambda_Q$ shift on  the RHS of
(\ref{ansatzNEW}) can be also linked to the KZ theory. To explain it,  we invoke the 
adiabatic-impulse approximation frequently employed in the context of the KZ theory of non-equilibrium 
time quenches.
This rather crude approximation assumes that 
in the crossover region between the two adiabatic regimes the state of the 
system is frozen, i.e., it does not change in the 
vicinity of  the critical point. 
Adopting such an approximation to the spatial quench, we would assume that in the crossover region, 
where the local homogeneous approximation breaks down, we have  
\be
\frac{dS^z}{dg}[g(x)]\approx \frac{dS^z_H}{dg}(g_c\pm\hat g)\approx\text{const} + \frac{\ln\lambda_Q}{2\pi}  \ \ \ \text{for} \  \ \
|x-x_c|\lesssim\hat x
\label{AIOH}
\ee
after using    (\ref{1_2pi}). 
While such a rough approximation says nothing about the location of the maximum 
of $dS^z/dg$, it properly predicts the  scaling of $dS^z/dg$  at the maximum. This
is seen by comparing (\ref{AIOH}) to  fit (\ref{fit_ln}), which we will
discuss below.
The adiabatic-impulse predictions always have to be taken with a grain of
salt \cite{FrancuzPRB2016}, which does not change the fact that they often times provide a
very reasonable lowest-order non-trivial  approximation (see e.g.
\cite{BDPRA2006,BDproceedings}).

Next, we proceed by subtracting $\ln\lambda_Q/2\pi$ from
$dS^z/dg$ followed by shifting and rescaling of the magnetic field 
\be
g\mapsto(g-1)\sqrt{\lambda_Q}.
\label{gma}
\ee
We  expect that after these transformations, data for different $\lambda_Q$'s should collapse on a single
curve, whose shape is given by the scaling function $f$. 
Fig. \ref{fig4}, however, shows that  the overlap of the
curves is not perfect, despite the fact that fairly high $\lambda_Q$'s are considered.

To look more quantitatively at ansatz (\ref{ansatzNEW}), we study $dS^z/dg$ at the maximum. 
Ansatz (\ref{ansatzNEW}) predicts that we should be expecting to find
\bee
\text{const} + \frac{\ln\lambda_Q}{2\pi}. 
\eee
In agreement with this prediction, we observe that the maximum of  $dS^z/dg$ can be very well
fitted with 
\be
A + B\ln\lambda_Q,
\label{fit_ln}
\ee
where $A\approx0.425$ and  $B\approx0.1584$ (Fig. \ref{fig5}a). Quite interestingly, the value of $B$  
very well  matches the expected $1/2\pi\approx0.159$. 

Turning now our attention  to the position $g^*_z$ of the maximum of $dS^z/dg$, we
fit it  with (\ref{fit}) getting $a\approx1.00001$, $b\approx-0.951$, and $c\approx-0.873$ (Fig. \ref{fig5}b).  
Similarly as in Sec. \ref{longitudinal_sec}, we obtain an excellent
extrapolation of the position of the critical point through the $a$  parameter.
The value of the exponent $c$, however, is surprising. Indeed, 
our expectation based on ansatz (\ref{ansatzNEW}) was that we would  get  $-1/(1+\nu)=-1/2$ instead.
It should be also noticed that the curve fitted in Fig. \ref{fig5}b does not match numerics for 
large enough $\lambda_Q$, which casts doubt on the power-law scaling of
$1-g^*_z$ with  $\lambda_Q$.

To resolve these issues, we propose to modify the renormalization procedure
taking full advantage of the knowledge of the exact expression for
the derivative of magnetization in the homogeneous system. 
To proceed, we write 
\be
\frac{dS^z_H}{dg}=\left.\frac{dS^z_H}{dg}\right|_\text{reg}-\frac{\chi_H(g)}{\pi}\ln|g-1|,
\label{szhex}
\ee
where the first (second) term on the RHS is  regular (singular)
at $g=1$. We mean by regularity  that the function itself and  all its derivatives are finite. The
$\chi_H(g)$ function is also regular at $g=1$, so that singularity  of 
$dS^z_H/dg$ comes solely from the logarithm. 
$\chi_H(g)$ and $\left.dS^z_H/dg\right|_\text{reg}$  
are given 
by  (\ref{chiH}) and (\ref{Szreg}), respectively.

Next, we define the observable $O$ in the inhomogeneous system as 
\bee
O=\B{\frac{dS^z}{dg}-\left.\frac{dS^z_H}{dg}\right|_\text{reg}}/\chi_H(g).
\eee
Its homogeneous system analogue $O_H$ is obtained  by replacing $dS^z/dg$ by $dS^z_H/dg$.
It is then immediately seen that $O_H$  transforms under the change of scale (\ref{gp_gc}) in the
same way as (\ref{OHmap}). Therefore, we again assume that
$O$ transforms in the same way as $O_H$ at least near the spatial critical point.
Repeating the above-performed steps, we arrive at the new scaling ansatz for
$dS^z/dg$, which we write in the following way
\be
f\B{\frac{g-g_c}{\hat g}}= \B{\frac{dS^z(g)}{dg}-\left.\frac{dS^z_H(g)}{dg}\right|_\text{reg}}/\chi_H(g)
-\frac{\ln\lambda_Q}{2\pi}.
\label{ansatzNEWNEW}
\ee
This ansatz reproduces ansatz (\ref{ansatzNEW}) when $\left.dS^z_H/dg\right|_\text{reg}$ and $\chi_H$
are evaluated at $g=1$ because  $\chi_H(1)=1$ and  $\left.dS^z_H/dg\right|_\text{reg}$
at $g=1$ can be absorbed  into the scaling  function  $f$. 
This follows from the fact that in the derivation of ansatz (\ref{ansatzNEW}),
we have assumed that $O_H$ is given by (\ref{1_2pi}).
Such an approximation is valid only extremely close to the critical point.
We overcome this limitation in the derivation of the ansatz
(\ref{ansatzNEWNEW}) using the exact solution for a homogeneous system.

\begin{figure}[t]
\includegraphics[width=\pref\columnwidth,clip=true]{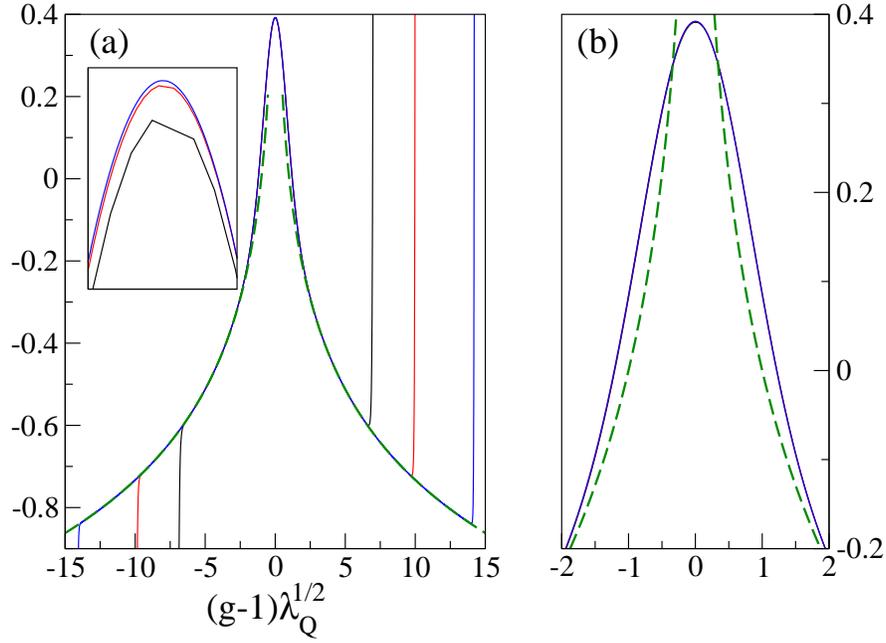}
\caption{Test of ansatz (\ref{ansatzNEWNEW}). 
Panel (a): solid lines depict RHS of  (\ref{ansatzNEWNEW}) for different $\lambda_Q$'s. The inset enlarges 
the extremum. The range of its horizontal and vertical axes is [-0.1,0.1] and
[0.387,0.3924], respectively.
Panel (b): the same as in the panel (a) but focused
around the maximum.   The green dashed line in both panels is given by (\ref{fxln}).
Results for $\lambda_Q$ equal to 
$1280$, $10240$, and $81910$ are depicted with black, red, and blue curves,
respectively. The curves overlap so well, that they are practically
indistinguishable in the central parts of both panels. The data for $dS^z/dg$,
used to prepare this plot, is the same as in  Fig. \ref{fig4}a. 
}
\label{fig6}
\end{figure}

If we now plot
the RHS of (\ref{ansatzNEWNEW}) for different $\lambda_Q$'s and rescale the
horizontal axis according to  (\ref{gma}), we find a wonderful collapse  of
different curves (Fig. \ref{fig6}). Its most remarkable feature is that it happens not only
close to the critical point, but also far away from it. Comparing 
Figs. \ref{fig4} and \ref{fig6}, we immediately see that ansatz (\ref{ansatzNEWNEW}) is significantly better
than ansatz (\ref{ansatzNEW}).

Next, we note that ansatz (\ref{ansatzNEWNEW}) reproduces  exactly the homogeneous result (\ref{szhex})  when 
\be
f(x)=-\frac{\ln|x|}{\pi}.
\label{fxln}
\ee 
Therefore, we expect that far away from the critical point, where the local
homogeneous approximation is supposed to work,  the scaling
function from (\ref{ansatzNEWNEW}) is accurately given by (\ref{fxln}).
Far away, in the light of the KZ theory, means that $|g-1|/\hat g\sim|g-1|\sqrt{\lambda_Q}\gg1$.
This is confirmed in Fig. \ref{fig6}a.

Finally, we discuss what ansatz (\ref{ansatzNEWNEW}) predicts for
the scaling of the  position of the maximum of $dS^z/dg$ with $\lambda_Q$. To proceed, we 
parametrize the scaling function around the maximum as
\bee
f(x)=ax^2+bx+c,
\eee
where $a$, $b$, and $c$ are some constants, and substitute the
expressions for $\chi_H$ (\ref{chiH}) and  $\left.dS_H^z/dg\right|_\text{reg}$ (\ref{Szreg}) into
(\ref{ansatzNEWNEW}) to obtain $dS^z/dg$ out of this equation. Next, we compute
from such an expression  $d^2S^z/dg^2$, expand it in a Taylor series around $g=1$ up to the linear in $g-1$ terms, 
and equal the resulting expression to zero. This gives us the  expression for the 
position $g^{**}_z$ of the extremum of $dS^z/dg$. The subsequent
discussion can be conveniently done by focusing on the distance $z^{**}$ of the maximum
from the critical point expressed in lattice spacing units:
\bee
z^{**}=(1-g_z^{**})\lambda_Q.
\eee
We get 
\be
z^{**}=\frac{4(
4 b\pi\sqrt{\lambda_Q} 
-3\ln\lambda_Q
-6c\pi 
-6 \ln8 
+14 )}{
32 a \pi 
-\frac{48b\pi}{\sqrt{\lambda_Q} }
+33\frac{\ln\lambda_Q}{\lambda_Q}
+\frac{66 c\pi 
+66\ln8 
-154}{\lambda_Q}
}.
\label{zstarstar}
\ee

\begin{figure}[t]
\includegraphics[width=\pref\columnwidth,clip=true]{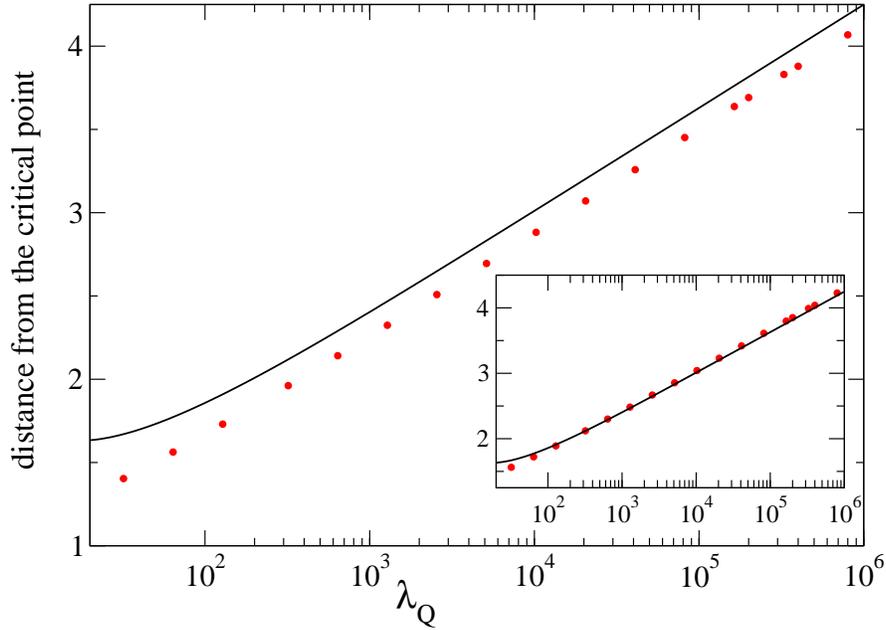}
\caption{Distance of the maximum of $dS^z/dg$ from the critical point in the
lattice spacing units. The red dots show numerics, i.e., 
$z^*=(1-g^*_z)\lambda_Q$. The black solid line shows $z^{**}$ for $a=-0.443$, $b=0$,
$c=0.392$; see   (\ref{zstarstar}). The inset shows the same as the main plot, but
after shifting  data for $z^*$ by $0.16$.
}
\label{fig7}
\end{figure}

The large $\lambda_Q$ scaling of  (\ref{zstarstar}) is $b\sqrt{\lambda_Q}/2a$, which is in
disagreement with numerical simulations, where $z^*=(1-g^*_z)\lambda_Q$ grows
approximately linearly in $\ln\lambda_Q$ (Fig. \ref{fig7}). Two remarks are in
order now.

First, the logarithmic growth of $z^*$ comes as a surprise if we look
at this result from the perspective of the KZ theory. The natural expectation
 would be that $z^*\sim \hat g\lambda_Q\sim \sqrt{\lambda_Q}$. It does
not, however, imply that the KZ scalings are absent in ansatz
(\ref{ansatzNEWNEW}). As we have explained above, they are present in the argument of
the scaling function $f$ and the term   $\sim\ln\lambda_Q$ on the RHS of
(\ref{ansatzNEWNEW}). In fact, given an excellent overlap of curves
in Fig. \ref{fig6}, the KZ scalings are visible not only close to the critical point, but
 everywhere in the system as long as we stay away from the points where the
 boundary conditions are imposed.

Second, the logarithmic scaling of $z^{**}$ is obtained from (\ref{zstarstar}) when $b=0$. Therefore, we set $b=0$ 
and get  $a$ and $c$ coefficients 
from the fit to the data rescaled in the same way as in Fig. \ref{fig6}. The
fits done for  $\lambda_Q\in(5\cdot10^3,4\cdot10^5)$ around the maximum  
yield  $a=-0.443(3)$ and $c=0.392(1)$. $z^{**}$ for these parameters is plotted in Fig. \ref{fig7}.
We see from this figure that $z^{*}$ and $z^{**}$ exhibit the same
logarithmic growth with $\lambda_Q$. They are, however, systematically shifted by about $0.16$ in lattice units (inset of Fig.
\ref{fig7}). 
We do not see an explanation for this curious shift of about $1/2\pi$; we
have extensively checked our numerical and analytical computations to make sure that they are technically correct. 

It should be stressed that data for $z^{**}$ comes from the
RG calculation that is  certainly  not expected to give definite predictions
with accuracy smaller than the distance between the lattice sites. Indeed, the RG
theory captures universal long-wavelength  behavior of the system rather than the
short-wavelength non-universal one. In fact, it is in
our opinion very interesting  that the discrepancy between the RG result and
the exact numerical one is so tiny.

\section{Summary}
\label{summary_sec}
We have comprehensively studied derivatives of  longitudinal and transverse magnetizations in the {\it inhomogeneous}
Ising model. We have shown that both these quantities exhibit an extremum that can be efficiently 
used for finding the position of the critical point of the homogeneous Ising chain. 

Properties of the extremum of the derivative of longitudinal magnetization 
are captured  by power law scalings  with the gradient of the external field. These
scaling relations encode  universal critical exponents. As a result, the correlation length
and longitudinal magnetization critical exponents, characterizing some singularities
of the homogeneous Ising chain, can be obtained from the studies of such an
extremum in the inhomogeneous system.

The scaling properties of the extremum of the derivative of transverse magnetization in the
inhomogeneous system are
rather complicated. Their proper description have required detailed knowledge of 
the exact solution in the homogeneous chain. The key difference between 
 derivatives of  longitudinal and transverse magnetizations is that 
in the homogeneous system the former exhibits a power law singularity while the
latter has a logarithmic singularity. It is a bit surprising that this  
makes such a  difference in the theoretical description of these derivatives in the  inhomogeneous system.  

Besides  exact diagonalization and  renormalization
group theory, one may apply the Kibble-Zurek approach to the
inhomogeneous Ising chain. Predictions of this intuitive  
approach are easily seen in the derivative of  longitudinal magnetization. 
Some of them are also present in the derivative of  transverse magnetization, but 
there is no straightforward way of extracting them. 

We believe that it would be interesting to extend these studies to other 
systems to gain  better understanding  of  quantum phase transitions
in inhomogeneous fields. Such studies should be useful in the context of  
 cold atom and ion experiments, where 
inhomogeneities of external fields are ubiquitous  and strongly-correlated  models describing
these systems are typically not exactly solvable  \cite{Lewenstein,BlochRMP2008}.
To experimentally study the critical points and exponents in these  systems, one may  try
to minimize  inhomogeneities disturbing their quantum phase transitions. 
This, however, may not always be feasible  (e.g.  the 
external trapping potential acting on cold  atoms and ions is present in
nearly all current experiments).
The other option is to control inhomogeneous fields and study
susceptibilities along the lines suggested in this work.

\section*{Acknowledgments}
 BD was supported by the Polish National Science Centre (NCN) grant
DEC-2013/09/B/ST3/00239. 
{M\L}    was supported by the European Research Council (ERC) Synergy
Grant UQUAM, the Austrian Science Fund through SFB FOQUS (FWF Project No.
F4016-N23), and EU FET Proactive Initiative SIQS. 
We  thank Marek Rams  for sharing with
us his detailed knowledge about the Bogolubov approach to the inhomogeneous Ising
model.

\appendix*
\section{Transverse magnetization in homogeneous Ising chain}
It is convenient to introduce the following notation
\bee
G=\frac{2\sqrt{g}}{1+g},  \  \ \tilde G=\sqrt{1-G^2}=\frac{|1-g|}{1+g},  \  \ 0\le G,\tilde G\le1.
\eee

The ground-state energy per lattice site in the thermodynamically-large homogeneous system is \cite{Lieb1961}
\bee
{\cal E}_H=-\frac{1}{\pi}\int_0^{\pi}dk\sqrt{1+g^2-2g\cos(k)}=-\frac{2}{\pi}(1+g)E(G),
\eee
where 
\bee
E(x)=\int_0^{\pi/2}dk\sqrt{1-x^2\sin^2(k)}
\eee
is the complete elliptic integral of the second kind \cite{Oldham}. 
Transverse magnetization is then given through the Feynman-Hellmann theorem,
$S^z_H=-d{\cal E}_H/dg$, which leads to (\ref{Sz_homo}).

The derivative of transverse magnetization then reads
\be
\frac{dS_H^z}{dg}=-\frac{d^2{\cal E}_H}{dg^2}=\frac{(1+g^2)K(G)-(1+g)^2E(G)}{g^2(1+g)\pi},
\label{dSzz}
\ee
where 
\bee
K(x)=\int_0^{\pi/2}\frac{dk}{\sqrt{1-x^2\sin^2(k)}}
\eee
is the complete elliptic integral of the first kind \cite{Oldham}.

Next, we need to factor out singular and regular at $g=1$ parts of elliptic integrals from  (\ref{dSzz}). Using identities 
from Sec. 61:6 of \cite{Oldham}, we get
\bee
\begin{aligned}
& \left.K(G)\right.=\left.K(G)\right|_\text{sing}+ \left.K(G)\right|_\text{reg},\\
& \left.K(G)\right|_\text{sing}=-\frac{2}{\pi}K(\tilde G)\ln|1-g|,
\  \ \left.K(G)\right|_\text{reg}=\frac{2}{\pi}K(\tilde G)\ln[4(1+g)] + 
2\sum_{j=1}^\infty\BB{\frac{(2j-1)!!}{(2j)!!}\tilde
G^j}^2\sum_{n=1}^{2j}\frac{(-1)^n}{n},\\
&\left.E(G)\right.=\left.E(G)\right|_\text{sing} + \left. E(G)\right|_\text{reg}, \\
&\left.E(G)\right|_\text{sing}=\frac{2}{\pi}\BB{E(\tilde G)-K(\tilde G)}\ln|1-g|,\\
&\left.E(G)\right|_\text{reg}=1+\frac{2}{\pi}\BB{K(\tilde G)-E(\tilde G)}\ln[4(1+g)] + 
2\sum_{j=1}^\infty\frac{(2j-1)!!(2j-3)!!}{(2j)!!(2j-2)!!}\tilde G^{2j}
\BB{\frac{1}{4j-8j^2}+\sum_{n=1}^{2j-2}\frac{(-1)^n}{n}}.\\
\end{aligned}
\eee
Note that $K(\tilde G)$ and $E(\tilde G)$ are regular at $g=1$ \cite{Oldham}.
These equations allow us to split $dS^z_H/dg$ into singular 
\be
\begin{aligned}
&\left.\frac{dS_H^z}{dg}\right|_\text{sing}=-\frac{\chi_H(g)}{\pi}\ln|1-g|,\\
&\left.\chi_H(g)\right.= \frac{2}{g^2(1+g)\pi} \BB{(1+g)^2E(\tilde G)-2gK(\tilde G)}=1+O(g-1),
\end{aligned}
\label{chiH}
\ee
and regular parts 
\be
\left.\frac{dS_H^z}{dg}\right|_\text{reg} =\frac{dS^z_H}{dg}-\left.\frac{dS^z_H}{dg}\right|_\text{sing}.
\label{Szreg}
\ee
The regular part is most conveniently obtained by replacing $K(G)$ and $E(G)$
in  (\ref{dSzz}) by $\left.K(G)\right|_\text{reg}$ and  $\left.E(G)\right|_\text{reg}$, respectively.


\begin{thebibliography}{36}
\expandafter\ifx\csname natexlab\endcsname\relax\def\natexlab#1{#1}\fi
\expandafter\ifx\csname bibnamefont\endcsname\relax
  \def\bibnamefont#1{#1}\fi
\expandafter\ifx\csname bibfnamefont\endcsname\relax
  \def\bibfnamefont#1{#1}\fi
\expandafter\ifx\csname citenamefont\endcsname\relax
  \def\citenamefont#1{#1}\fi
\expandafter\ifx\csname url\endcsname\relax
  \def\url#1{\texttt{#1}}\fi
\expandafter\ifx\csname urlprefix\endcsname\relax\def\urlprefix{URL }\fi
\providecommand{\bibinfo}[2]{#2}
\providecommand{\eprint}[2][]{\url{#2}}

\bibitem[{Kib()}]{Kibble1976}
\bibinfo{note}{T. W. B. Kibble, J. Phys. A {\bf 9}, 1387 (1976); Phys. Rep.
  {\bf 67},183 (1980).}

\bibitem[{Zur()}]{Zurek1985}
\bibinfo{note}{W. H. Zurek, Nature (London) {\bf 317}, 505 (1985); W.H. Zurek,
  Phys. Rep. {\bf 276}, 177 (1996).}

\bibitem[{KZc()}]{KZclassRev}
\bibinfo{note}{T. W. B. Kibble, Phys. Today {\bf 60}, 47 (2007); A. del Campo,
  T. W. B. Kibble, and W. H. Zurek, J. Phys.: Condens. Matter {\bf 25}, 404210
  (2013); A. del Campo and W. H. Zurek, Int. J. Mod. Phys. A {\bf 29}, 1430018
  (2014).}

\bibitem[{BDP({\natexlab{a}})}]{BDPRL2005}
\bibinfo{note}{B. Damski, Phys. Rev. Lett. {\bf 95}, 035701 (2005).}

\bibitem[{Dor({\natexlab{a}})}]{DornerPRL2005}
\bibinfo{note}{W. H. Zurek, U. Dorner, and P. Zoller, Phys. Rev. Lett. {\bf
  95}, 105701 (2005).}

\bibitem[{Jac({\natexlab{a}})}]{JacekPRL2005}
\bibinfo{note}{J. Dziarmaga, Phys. Rev. Lett. {\bf 95}, 245701 (2005).}

\bibitem[{Pol({\natexlab{a}})}]{PolkovnikovPRB2005}
\bibinfo{note}{A. Polkovnikov, Phys. Rev. B {\bf 72}, 161201(R) (2005).}

\bibitem[{Jac({\natexlab{b}})}]{JacekAdv2010}
\bibinfo{note}{J. Dziarmaga, Adv. Phys. 59, 1063 (2010).}

\bibitem[{Pol({\natexlab{b}})}]{PolkovnikovRMP2011}
\bibinfo{note}{A. Polkovnikov, K. Sengupta, A. Silva, and M. Vengalattore, Rev.
  Mod. Phys. {\bf 83}, 863 (2011).}

\bibitem[{Dut()}]{Dutta2015}
\bibinfo{note}{A. Dutta, G. Aeppli, B. K. Chakrabarti, U. Divakaran, T. F.
  Rosenbaum, and D. Sen, {\it Quantum Phase Transitions in Transverse Field
  Spin Models: From Statistical Physics to Quantum Information} (Cambridge
  University Press, 2015); arXiv:1012.0653.}

\bibitem[{Tur({\natexlab{a}})}]{TurbanJPA2007}
\bibinfo{note}{T. Platini, D. Karevski, and L. Turban, J. Phys. A: Math. Theor.
  {\bf 40}, 1467 (2007).}

\bibitem[{Dor({\natexlab{b}})}]{DornerPRS2008}
\bibinfo{note}{W. H. Zurek and U. Dorner, Phil. Trans. R. Soc. A{\bf 366}, 2953
  (2008).}

\bibitem[{Tur({\natexlab{b}})}]{TurbanJStat2009}
\bibinfo{note}{M. Collura, D. Karevski, and L. Turban, J. Stat. Mech. (2009)
  P08007.}

\bibitem[{BDN()}]{BDNJP2009}
\bibinfo{note}{B. Damski and W. H. Zurek, New J. Phys. {\bf 11}, 063014
  (2009).}

\bibitem[{Cam()}]{CampostriniPRA2010}
\bibinfo{note}{M. Campostrini and E. Vicari, Phys. Rev. A {\bf 81}, 023606
  (2010).}

\bibitem[{Suz()}]{SuzukiPRB2015}
\bibinfo{note}{S. Suzuki and A. Dutta, Phys. Rev. B {\bf 92}, 064419 (2015).}

\bibitem[{Jac({\natexlab{c}})}]{JacekMarekNJP2010}
\bibinfo{note}{J. Dziarmaga and M. M. Rams, New J. Phys. {\bf 12} 055007
  (2010); New J. Phys. {\bf 12}, 103002 (2010).}

\bibitem[{Col()}]{ColluraPRA2011}
\bibinfo{note}{M. Collura and D. Karevski, Phys. Rev. Lett. {\bf 104}, 200601
  (2010); Phys. Rev. A {\bf 83}, 023603 (2011).}

\bibitem[{Mar()}]{MarekAdolfoNJP2016}
\bibinfo{note}{M. M. Rams, M. Mohseni, and A. del Campo, New J. Phys. {\bf 18},
  123034 (2016).}

\bibitem[{Lew()}]{Lewenstein}
\bibinfo{note}{M. Lewenstein, A. Sanpera, and V. Ahufinger, {\it Ultracold
  Atoms in Optical Lattices: Simulating Quantum Many-Body Systems} (Oxford
  University Press, Oxford, UK, 2012).}

\bibitem[{Blo()}]{BlochRMP2008}
\bibinfo{note}{I. Bloch, J. Dalibard, and W. Zwerger, Rev. Mod. Phys. {\bf 80},
  885 (2008).}

\bibitem[{Car({\natexlab{a}})}]{Cardy1}
\bibinfo{note}{J. L. Cardy, ed., {\it Finite-Size Scaling} (North-Holland,
  Amsterdam, 1988).}

\bibitem[{Kol()}]{KolodrubetzPRL2012}
\bibinfo{note}{M. Kolodrubetz, B. K. Clark, and D. A. Huse, Phys. Rev. Lett.
  {\bf 109}, 015701 (2012).}

\bibitem[{Son()}]{SondhiPRB2012}
\bibinfo{note}{A. Chandran, A. Erez, S. S. Gubser, and S. L. Sondhi, Phys. Rev.
  B {\bf 86}, 064304 (2012).}

\bibitem[{Fra()}]{FrancuzPRB2016}
\bibinfo{note}{A. Francuz, J. Dziarmaga, B. Gardas, and W. H. Zurek, Phys. Rev.
  B {\bf 93}, 075134 (2016).}

\bibitem[{BDP({\natexlab{b}})}]{BDPRL2010}
\bibinfo{note}{B. Damski and W. H. Zurek, Phys. Rev. Lett. {\bf 104}, 160404
  (2010).}

\bibitem[{Lie()}]{Lieb1961}
\bibinfo{note}{E. Lieb, T. Schultz, and D. Mattis, Ann. Phys. (N.Y.) {\bf 16},
  407 (1961).}

\bibitem[{Pfe()}]{Pfeuty}
\bibinfo{note}{P. Pfeuty, Ann. Phys. {\bf 57}, 79 (1970).}

\bibitem[{Bar()}]{Barouch1971}
\bibinfo{note}{E. Barouch and B. M. McCoy, Phys. Rev. A {\bf 3}, 786 (1971).}

\bibitem[{You()}]{YoungPRB1997}
\bibinfo{note}{A. P. Young, Phys. Rev. B {\bf 56}, 11691 (1997).}

\bibitem[{Wic()}]{Wick}
\bibinfo{note}{G. F. Giuliani and G. Vignale, {\it Quantum Theory of the
  Electron Liquid} (Cambridge University Press, 2005).}

\bibitem[{Mat()}]{Mathematica}
\bibinfo{note}{Wolfram Research, Inc., Mathematica, Version 11.0, Champaign, IL
  (2016).}

\bibitem[{Car({\natexlab{b}})}]{Cardy2}
\bibinfo{note}{J. Cardy, {\it Scaling and Renormalization in Statistical
  Physics} (Cambridge University Press, Cambridge, 2002).}

\bibitem[{BDP({\natexlab{c}})}]{BDPRA2006}
\bibinfo{note}{B. Damski and W. H. Zurek, Phys. Rev. A {\bf 73}, 063405
  (2006).}

\bibitem[{BDp()}]{BDproceedings}
\bibinfo{note}{B. Damski, Fidelity approach to quantum phase transitions in
  quantum Ising model, in {\it Quantum Criticality in Con- densed Matter:
  Phenomena, Materials and Ideas in Theory and Experiment}, edited by J.
  Jedrzejewski (World Scientific, Singapore, 2015), pp. 159–182;
  arXiv:1509.03051.}

\bibitem[{Old()}]{Oldham}
\bibinfo{note}{K. Oldham, J. Myland, and J. Spanier, {\it An Atlas of
  Functions}, 2nd ed. (Springer, 2009).}

\end{thebibliography}

\end{document}